\documentclass[a4paper]{jpconf}
\usepackage{graphicx}
\usepackage{amsbsy}
\usepackage{amssymb}

\newcommand{\I}{\mathrm{i}} 
\newcommand{\D}{\mathrm{d}} 
\newcommand{\E}{\mathrm{e}} 

\begin{document}
\title{Coordinate-space representation of a charged scalar \\ particle propagator in a constant magnetic field \\ expanded as a sum over the Landau levels}

\author{S N Iablokov $^{[1,2] \, *}$ and A V Kuznetsov $^{[1]}$}
\address{$^{[1]}$ P.G. Demidov Yaroslavl State University, Yaroslavl, Russia} 
\address{$^{[2]}$ A.A. Kharkevich Institute for Information Transmission Problems, Moscow, Russia}

\ead{$^{*}$physics@iablokov.ru}

\begin{abstract}
A coordinate-space representation for a charged scalar particle propagator in a constant magnetic field was obtained as a series over the Landau levels. Using the recently developed modified Fock--Schwinger method, an intermediate expression was constructed and symmetrized, thus, allowing for factorization of the series terms into two factors. The first one, a sum of Bessel functions, depends on time and $z$-coordinate, where the $z$-axis is chosen to be a direction of the magnetic field, and has a structure similar to the propagator of a free field. The second one, a product of a Laguerre polynomial and a damping exponential, depends on $x,y$-coordinates, which form a plane perpendicular to the direction of the magnetic field, and ensures the localized propagation in the $x,y$-plane.
\end{abstract}

\section{Introduction}
There exist different representations of charged particle propagators in a constant magnetic field. Among the most useful are the Fock-Schwinger proper-time representation \cite{Schwinger,Itzykson_1980}, both in the coordinate and momentum spaces, and the momentum-space representation as an expansion over the Landau levels \cite{KM_Book_2013}. In this study we derive the missing coordinate-space representation for the propagator of a charged scalar particle as a series over the Landau levels. As a starting point we choose our recently published modification \cite{MFS_2018,MFS_2020} of the Fock-Schwinger (FS) proper time method, which allows to rapidly obtain an intermediate expression as a partial Fourier image. Using standard integration techniques, we evaluate the remaining Fourier integrals and obtain the coordinate-space expression of the propagator expanded as a sum over the Landau levels.

\section{Outline of the modified Fock---Schwinger approach}

There are many ways to obtain propagators of charged particles in a constant magnetic field, one of which consists in solving the corresponding propagator equation provided by the path integral formalism:
\begin{equation}
\label{eq_hg_delta}
H(\partial_X, X) \, G(X, X') = \delta^{(4)} (X-X') \, ,
\end{equation}
where $X^\mu = (t, x, y, z)$ and $X'^\mu = (t', x', y', z')$ are the space-time 4-vectors. In this study we apply {\it{modified}} Fock-Schwinger (MFS) method to get the solution of Eq.~(\ref{eq_hg_delta}). First, $G(X, X')$ is represented as an integral:
\begin{equation}
\label{q}
G(X, X') = -\I \int_{-\infty}^{0} \D \tau \, U(X,X';\tau) \, ,
\end{equation}
where $\tau$ is called the proper time. The operator $U(X,X';\tau)$ could be considered as some kind of an evolution operator which satisfies a Schr{\"o}dinger-type equation
\begin{equation}
\label{eq_u_shroedinger}
\I \, \partial_\tau \, U(X, X'; \tau) = H(\partial_X, X) \, U(X, X'; \tau) \,. 
\end{equation}
Taking the appropriate boundary conditions
\begin{eqnarray}
\label{u_boundary}
U(X,X';-\infty) = 0 \quad \quad \quad U(X,X';0) = \delta^{(4)} (X-X') \, ,
\end{eqnarray}
one could obtain the following expression:
\begin{eqnarray}
\label{u_exp_delta}
U(X, X'; \tau) &=& \E^{ -\I \tau \left[ \, H(\partial_X, X) + \I \varepsilon \right] } \, \delta^{(4)} (X-X') \, .
\end{eqnarray}
Contrary to the original FS approach, the MFS method consists in the direct evaluation of the exponential operator action on the $\delta$-function. In order to do so, an appropriate representation of the $\delta$-function as an integral and/or series should be chosen:
\begin{equation}
\label{eq_delta_representation}
\delta^{(4)} (X-X') = \sum_{\lambda} \int \psi_\lambda(X) \psi_\lambda(X') \, ,
\end{equation}
where $\psi_\lambda(X)$ is an eigenvector of the $H$ operator:
\begin{equation}
\label{eigen_h}
H(\partial_X, X) \psi_\lambda(X) = H(\lambda) \psi_\lambda(X) \, .
\end{equation}
This leads to the following simplification: 
\begin{equation}
\label{eq_exp_tau}
G(X, X') = -\I \int_{-\infty}^{0} \D \tau \, \sum_{\lambda} \int \E^{ -\I  \tau \, \left[ H(\lambda) + \I \varepsilon \right] }  \psi_\lambda(X) \psi_\lambda(X') \, .
\end{equation}
Finally, the exponential part is integrated out:
\begin{equation}
\label{eq_7b}
G(X, X') =  \sum_{\lambda} \int \frac{\psi_\lambda(X) \psi_\lambda(X')}{ H(\lambda) + \I \varepsilon} \, .
\end{equation}

The resulting expression is an integral/series over the continuous/discrete quantum numbers, and does not contain the proper time parameter. 
The MFS method is one of the fastest ways to obtain propagators, especially in the external fields for particles with spin. We will apply it in the next section to get an intermediate form of the propagator which will serve as a starting point for the calculation of the coordinate-space representation expanded in a series over the Landau levels.

\section{Calculation of a charged scalar particle propagator}

We apply the MFS method to solve the propagator equation for a charged scalar particle:
\begin{equation}
\label{eq_7b}
\bigg[ \left( \I \partial_\mu - eQA_\mu \right)^2 - m^2 \bigg] \, G(X, X') = \delta^{(4)} (X-X') \, .
\end{equation}
For convenience, we direct magnetic field $\vec{B}$ along the $z$-axis and make calculations in the Landau gauge: $A^\mu = (0, 0, Bx, 0)$. With this choice of gauge, the $H$ operator obtains the form:
\begin{equation}
\label{eq_7g}
H  = \left( \I \partial_\mu - eQA_\mu \right)^2 - m^2 =  p^2_\parallel - m^2 + \beta(\partial^2_\eta -\eta^2) \, .
\end{equation}
The substitutions $-\I \partial_0 \to p_0$, $\I \partial_y \to p_y$ and $\I \partial_z \to p_z$ are justified by the appropriate representation of the $\delta$-function:
\begin{equation}
\label{delta_repr}
\delta^{(4)} (Z) = \sqrt{\beta} \sum_{n=0}^\infty \int \frac{\D ^3 p_{\,\shortparallel, y}}{(2\pi)^3} \,
\E^{{-\I \left( pZ \right)}_{\shortparallel, y}} V_n V_n ' \, .
\end{equation}
Here, $\beta = eB$, $\eta = \sqrt{\beta} \, \big( x - Q\frac{p_y}{\beta} \big)$, $\eta ' = \sqrt{\beta} \, \big( x' - Q\frac{p_y}{\beta} \big)$, $Z^\mu = X^\mu - X'^{\mu}$ and $||$ stands for $t,z$-coordinates. The functions $V_n = V_n(\eta)$ and $V_n^{'} = V_n(\eta ')$ are the the $n$-th level quantum harmonic oscillator (QHO) eigenfunctions. Using the formula (8) and the QHO eigenvalue equation 
\begin{equation}
\label{qho_eigen}
(\partial^2_\eta -\eta^2)V_n = -(2n+1)V_n \, ,
\end{equation}
one obtains  the following representation of the propagator:
\begin{eqnarray}
\label{eq_result_exp_h}
G(X, X') = - \I \sqrt{\beta} \sum_{n=0}^\infty \int \frac{\D ^3 p_{\,\shortparallel, y}}{(2\pi)^3} \, \int_{-\infty}^{0} \D \tau \,
\E^{-\I \tau \left( p^2_\parallel - M_n^2 + \I \varepsilon \right)} \E^{{-\I \left( pZ \right)}_{\shortparallel, y}} V_n V_n ' \, ,
\end{eqnarray}
where $M_n^2 = m^2 + (2n+1)\beta$. A straightforward evaluation of the integral over $\tau$ leads to:
\begin{eqnarray}
\label{g_py_repr}
G(X, X') = \sqrt{\beta} \sum_{n=0}^\infty \int \frac{\D ^3 p_{\,\shortparallel, y}}{(2\pi)^3}  \frac{\E^{{- \I \left( pZ \right)}_{\shortparallel, y}}}{p^2_\parallel - M_n^2 + \I \varepsilon} V_n V_n ' \, .
\end{eqnarray}
This form of the propagator, however, is not invariant with respect to rotations in the $x,y$-plane and, thus, does not exhibit the internal symmetry of the problem. In order to symmetrize it, we use the following formula from Ref. \cite{GR}:
\begin{eqnarray}
\label{eq_7.4}
I_{n,n'} = \int_{-\infty}^\infty \D u \, \E^{-u^2} H_n (u+a) H_{n'} (u+b) = 2^{n'} \sqrt{\pi} \, n! \, b^{n'-n} L^{(n'-n)}_{n}(-2ab) \quad (n' \ge n) \,,
\end{eqnarray}
where $L_n^{(m)}$ are the associated Laguerre polynomials. 

Making an appropriate change of variables in Eq.~(\ref{g_py_repr}) (see, e.g., Ref. \cite{MFS_2020}) and integrating over $p_y$ leads to the following symmetrized representation:
\begin{eqnarray}
G(X, X') = \frac{\beta}{2 \pi} \E^{\I \Phi} \sum_{n=0}^\infty \, L_n \bigg(\frac{\beta Z^2_\perp}{2}\bigg) \, \E^{-\beta Z^2_\perp / 4} \, \int \frac{\D^2 p_{\,\shortparallel}}{(2\pi)^2} \frac{\E^{{- \I \left( pZ \right)}_{\parallel}}}{p^2_\parallel - M_n^2 + \I \varepsilon} \, .
\end{eqnarray}
Here, $\perp$ stands for the $x,y$-coordinates. We also note an overall non-invariant phase factor $e^{\I \Phi}$ where $\Phi(X,X') = \frac{Q\beta}{2} (x+x')(y-y')$. Aside from this phase, the summation terms decompose into 2 factors. The first one depends only on the $x,y$-coordinates and is invariant with respect to rotations in the $x,y$-plane which is perpendicular to the direction of the magnetic field. The second factor (the Fourier integral) 
\begin{eqnarray}
J_{\parallel} = \int \frac{\D^2 p_{\,\shortparallel}}{(2\pi)^2} \frac{\E^{{- \I \left( pZ \right)}_{\parallel}}}{p^2_\parallel - M_n^2 + \I \varepsilon}
\end{eqnarray}
depends only on the $t,z$-coordinates and allows for further simplification. The integration proceeds much like in the case of a free field \cite{BS} and relies on the following integral 
identity from Ref.~\cite{GR}:
\begin{eqnarray}
\int_{0}^{\infty} \D u \, \cos(bu) \frac{\E^{-a\sqrt{c^2+u^2}}}{\sqrt{c^2 + u^2}} = K_0 \bigg( c\sqrt{a^2+b^2} \bigg) \, , 
\end{eqnarray}
where $K_0$ is the modified Bessel function of the second kind. The resulting expression can be written as a distribution (generalized function):
\begin{eqnarray}
J_{\parallel} = \lim_{\epsilon \to 0^{+}} \bigg[ \frac{-\I}{2\pi } K_0 \bigg( M_n \sqrt{-Z_\parallel^2 + \I \epsilon} \bigg) \bigg] \, .
\end{eqnarray}
We keep this expression for $Z_\parallel^2 < 0$. For the case when $Z_\parallel^2 > 0$, we transform it by applying another useful identity from Ref. \cite{GR}:
\begin{eqnarray}
K_\nu(z) = -\frac{\I\pi}{2} \, \E^{-\I\pi\nu/2} H_{-\nu}^{(2)}\bigg( z \E^{-\I \pi/2} \bigg) \, ,
\end{eqnarray}
where $H_\nu^{(2)}$ are the Hankel functions of the second kind.
The final expression for the coordinate-space representation of the propagator reads:
\begin{eqnarray}
G(X, X') &=& \frac{-\I\beta}{4\pi^2} \, \E^{\I \Phi} \lim_{\epsilon \to 0^{+}} \sum_{n=0}^{\infty} \, L_n \bigg(\frac{\beta Z^2_\perp}{2}\bigg) \, \E^{-\beta Z^2_\perp / 4} \, \times
\\
\nonumber
&& \bigg\{ K_0 \bigg( M_n \sqrt{-Z_\parallel^2 + \I \epsilon} \bigg) \, \theta(-Z_\parallel^2) - \frac{\I\pi}{2} \, H_0^{(2)} \bigg( M_n \sqrt{Z_\parallel^2 - \I \epsilon} \bigg) \, \theta(Z_\parallel^2) \bigg\} \, .
\end{eqnarray}
%

\section{Conclusion}
Using the modified Fock-Schwinger method we were able to rapidly obtain the symmetrized representation of the propagator, which allowed for a straightforward evaluation of the remaining Fourier integral. The final expression in the coordinate space written as a series over the Landau levels decomposed into several parts. Each term in the series corresponds to a particular Landau level and consists of two factors. The first factor depends only on the coordinates in the $x,y$-plane which is perpendicular to the direction of the magnetic field, and is invariant with respect to rotations in this plane. The damping exponential ensures that the propagation in the $x,y$-plane is localized. The second factor describes propagation in the $t,z$-plane and is similar to the case of free field, however, in Minkowski 2-space instead of 4-space. We also note the non-invariant overall phase factor $e^{\I\Phi}$ which naturally arises when considering propagators of charged particles in a constant magnetic field.

\ack
The reported study was funded by RFBR, project number 19-32-90137.

\smallskip

\end{document}